\documentclass[fleqn,usenatbib]{mnras}

\usepackage{newtxtext,newtxmath}
\usepackage[T1]{fontenc}
\usepackage{rotating}
\usepackage{xcolor}
\usepackage{fontspec}
\usepackage{xeCJK}
\DeclareRobustCommand{\VAN}[3]{#2}
\let\VANthebibliography\thebibliography
\def\thebibliography{\DeclareRobustCommand{\VAN}[3]{##3}\VANthebibliography}

\usepackage{graphicx}	% Including figure files
\usepackage{amsmath}% Advanced maths commands
\newcommand{\rev}[1]{#1}
\newcommand{\revt}[1]{\textcolor{black}{#1}}

\title[Searching FRBs from M49 GCs with FAST]{
A search for Fast Radio Bursts from globular clusters in M49 with FAST
}

\author[Ho et al.]{
Simon C.-C. Ho (何建璋),$^{1,2,3}$\thanks{E-mail: Simon.Ho@anu.edu.au}
Chris Flynn,$^{2,3}$
Matthew Bailes,$^{2,3}$
Emma Carli,$^{2,3}$
Lei Zhang (张蕾),$^{2,3,4}$
\newauthor
Manisha Caleb,$^{5,3}$
Kenneth C.~Freeman,$^{1}$
Tetsuya Hashimoto,$^{6}$
Tomotsugu Goto$^{7,8}$
and James O. Chibueze$^{9,10}$
\\
$^{1}$Research School of Astronomy and Astrophysics, The Australian National University, Canberra, ACT 2611, Australia\\
$^{2}$Centre for Astrophysics and Supercomputing, Swinburne University of Technology, Hawthorn, VIC 3122, Australia\\
$^{3}$OzGrav: The Australian Research Council Centre of Excellence for Gravitational Wave Discovery, Hawthorn, VIC 3122, Australia\\
$^{4}$State Key Laboratory of Radio Astronomy and Technology, National Astronomical Observatories, Chinese Academy of Sciences, Beijing 100101, China\\
$^{5}$Sydney Institute for Astronomy, School of Physics, The University of Sydney, Sydney, NSW 2006, Australia\\
$^{6}$Department of Physics, National Chung Hsing University, 145 Xingda Rd., South Dist., Taichung 40227, Taiwan\\
$^{7}$Department of Physics, National Tsing Hua University, 101, Section 2. Kuang-Fu Road, Hsinchu, 30013, Taiwan\\
$^{8}$Institute of Astronomy, National Tsing Hua University, 101, Section 2. Kuang-Fu Road, Hsinchu, 30013, Taiwan\\
$^{9}$UNISA Centre for Astrophysics and Space Sciences (U-CASS), College of Science, Engineering and Technology, University of South Africa, Cnr Christian\\ de Wet Rd and Pioneer Avenue, Florida Park, 1709, Roodepoort, South Africa.\\
$^{10}$Department of Physics and Astronomy, Faculty of Physical Sciences, University of Nigeria, Carver Building, 1 University Road, Nsukka 410001, Nigeria.
}

\pubyear{2026}

\begin{document}
%\begin{CJK*}{UTF8}{gbsn}
\label{firstpage}
\pagerange{\pageref{firstpage}--\pageref{lastpage}}
\maketitle

\begin{abstract}

The origins of fast radio bursts (FRBs) remain uncertain, although magnetars are a leading progenitor candidate. Because magnetars are thought to form primarily through core-collapse supernovae in young stellar populations, the discovery of FRB\,20200120E in a globular cluster (GC) in the nearby galaxy M81 was unexpected given the ancient stellar populations of GCs. Expanding the sample of FRBs localised to nearby galaxies is therefore essential for testing FRB formation channels in old stellar environments. M49 (NGC\,4472) is a nearby ($\approx17$\,Mpc), radio-quiet giant elliptical galaxy in the Virgo cluster hosting an extensive GC system ($\approx7000$\,GCs), making it an ideal target for GC FRB searches. We conducted a 9-hour SnapShotCal observation of M49 using the Five-hundred-meter Aperture Spherical Telescope (FAST) 19-beam receiver, covering approximately 4230\,GCs \rev{($T_{\rm eff}=\revt{2.1}$\,hr per GC)}, and performed a comprehensive single-pulse search with \textsc{TransientX} over a dispersion-measure range of $0$--$5000\,\mathrm{pc\,cm^{-3}}$. No unambiguous astrophysical FRBs were detected. The most significant trigger reached a post-processed signal-to-noise ratio of $8.6\,\sigma$ at a dispersion measure of $412.2\,\mathrm{pc\,cm^{-3}}$, but is statistically consistent with thermal noise after accounting for the false-alarm rate. Using the radiometer equation, we derive a \rev{beam-averaged} peak flux-density sensitivity of \rev{$\approx16.5$\,mJy} (\rev{corresponding to a} fluence limit of \rev{$\approx16.5\,\mathrm{mJy\,ms}$} for a 1\,ms burst) and place an upper limit on the FRB occurrence rate of $\revt{4.7}\times10^{-4}\,\mathrm{FRB\,GC^{-1}\,hr^{-1}}$. \rev{Our non-detection implies that no bright bursts were observed above this fluence threshold during the observing window. The derived rate limit is therefore sensitivity-limited and applies only to bursts above $\approx16.5\,\mathrm{mJy\,ms}$ (for a 1\,ms burst), constraining only the detectable (high-fluence) portion of the GC FRB population.}

\end{abstract}

%%%%%%%%%%%%%%%%%%%%%%%%%%%%%%%%%%%%%%%%%%%%%%%%%%

%%%%%%%%%%%%%%%%% BODY OF PAPER %%%%%%%%%%%%%%%%%%

\section{Introduction} \label{sec:M49:intro}
Fast radio bursts (FRBs) are millisecond radio flashes of unknown origin. 
Since the discovery of FRBs more than a decade ago \citep{Lorimer2007}, and their association with host galaxies at cosmological distances \citep{Bhandari2022, Gordon2023}, a large number of search and follow-up programs have been undertaken globally \citep[e.g.,][]{Macquart2010, CHIME2018}.
However, with thousands of events detected, the progenitors of FRBs remain under debate \citep{Petroff2022, Zhang2023, Lorimer2024}. In 2020, a low-luminosity FRB repeater FRB\,200428, was found to be associated with a Galactic magnetar, SGR\,1935+2154 \citep{Bochenek2020, CHIME2020}, providing compelling evidence that magnetars are capable of producing FRB-like emission. These strongly magnetic and young neutron stars are usually expected to be found inside star-forming regions within galaxies, possibly within supernova remnants \citep{Vink2006, Kaspi2017}. 

\citet{Bhardwaj2021} and \citet{Kirsten2022} discovered FRB\,20200120E, a repeating FRB localised to a globular cluster (GC) associated with the nearby spiral galaxy M81, using the Canadian Hydrogen Intensity Mapping Experiment (CHIME) and the European Very Long Baseline Interferometry (VLBI) network (EVN). The host GC exhibits properties typical of ancient, metal-poor clusters, with an estimated metallicity of $\mathrm{[Fe/H]} \approx -1.5$ and an age of $\gtrsim 10$~Gyr \citep{Kirsten2022}. Metal-poor GCs like this are common in both spiral and elliptical galaxies, especially in massive ellipticals that host large populations of old clusters. This discovery challenges the conventional magnetar model for FRB origins, suggesting that at least some FRBs may arise from older stellar populations. Supporting this unusual behaviour, \citet{Nimmo2023} reported a `burst storm' from FRB 20200120E, detecting 53 bursts with fluences down to 0.04 Jy ms within just 40 minutes, indicative of episodes of intense activity.
The presence of an FRB in a GC is particularly intriguing because GCs host predominantly old stellar populations, whereas magnetars, the leading FRB progenitor candidates, are thought to be young neutron stars formed via core-collapse supernovae. This apparent discrepancy motivates alternative formation channels in such environments. \citet{Kirsten2022}, \citet{Kremer2023}, and \citet{Lu2022} proposed that highly magnetised neutron stars in GCs could form via accretion-induced collapse (AIC) of a white dwarf or the merger of massive white dwarf binaries, processes expected to be common in dense cluster environments \citep[e.g.,][]{Kremer2021}. An alternative possibility is that GC FRBs represent extreme manifestations of giant radio pulse emission from energetic millisecond pulsars \citep{Cordes2016}. GCs host abundant recycled pulsars, some of which exhibit prolific and luminous GP activity \citep[e.g.,][]{Ho2025}. If similar emission mechanisms operate at even higher energies, FRB~20200120E may represent the extreme tail of magnetospheric emission from an old neutron star in a dense stellar system \citep{Lyutikov2016}.

Expanding the sample of FRBs associated with GCs is critical for constraining their origins. To date, however, FRB\,20200120E in M81 remains the only FRB confirmed to be hosted by a GC. This may reflect the fact that such sources represent only a small sub-population of the overall FRB population \citep{Bhardwaj2024, Gordon2025}. Moreover, because GCs are intrinsically faint in the optical, robust host associations require FRBs to be sufficiently nearby. This observational limitation is illustrated by FRB\,20240209A, recently reported by \citet{Eftekhari2025} and \citet{Shah2025}, which was localised using CHIME outriggers \citep{Lanman2024} to the outskirts of a massive and quiescent elliptical galaxy. A GC origin is strongly favoured due to the FRB's large projected offset from its host galaxy; however, its relatively high redshift ($z = 0.1384$) makes optical detection of an associated GC challenging, and the GC association remains unconfirmed. Indeed, FRB\,20200120E is currently the nearest extragalactic FRB with precise sub-arcsecond localisation. These considerations suggest that targeted searches of nearby galaxies with rich GC systems provide the best opportunity to uncover further FRB-GC associations.

In this paper, we report a 9-hour observation of M49 with the FAST aimed at searching for FRBs originating from its GC systems. Owing to its unprecedented sensitivity (SEFD $\approx$ 2\,Jy at L-band), FAST provides a uniquely powerful probe of faint FRB emission from nearby GC systems. Consequently, even a non-detection places particularly stringent upper limits on the FRB burst rate and energetics in M49. We describe target selection in Section \ref{sec:M49sampleselection} and the observation in Section \ref{sec:M49observation}. The data pre-processing and single-pulse search are described in Section \ref{sec:M49search}. The results are presented in Section \ref{sec:M49results}. The discussion is in Section \ref{sec:M49discussion}. Lastly, we have our conclusions in Section \ref{sec:M49conclusion}.

\section{Target selection} \label{sec:M49sampleselection}
Several nearby galaxies are known to host hundreds to thousands of GCs, and thus constitute promising targets for such efforts \citep{Harris2013}. Searches have been made for giant radio pulses from extragalactic targets with GCs such as NGC\,253, NGC\,300, NGC\,6300, NGC\,7793, and the Fornax galaxy by using a single beam from the Parkes (Murriyang) Telescope for 30 minutes each, but none were found \citep{McLaughlin2003}. \citet{Hurter2024} carried out pulsar, magnetar and FRB searches in the nearer (3.5\,Mpc) spiral Sculptor Galaxy, NGC\,253, with the MeerKAT telescope in coherent mode for $\approx 4$ hr. However, they did not detect any pulsars or single pulses but derived a single pulse peak flux density limit of 62\,mJy at 1284 MHz. They suggested future observations covering a longer integration time may uncover pulsars and FRBs in this and other nearby galaxies. \citet{Kremer2023} estimated that a radio survey lasting roughly 100 hours directed at several massive galaxies located within $\approx15$\,Mpc, with a fluence threshold of 0.1\,Jy ms, is expected to detect on the order of 10 bursts, some of which may be localised to potential host star clusters similar to the M81 source, based on the observed characteristics of the M81 FRB. 
They estimate the FRB rate for $\approx 40$ nearby galaxies with distance $\leqq$ 20\,Mpc. They scale the detection rate of repeat bursts in the M81 FRB from CHIME and provide 90\% confidence range (which is inferred from adopting a Poisson distribution from \citealt{Lu2022}). Seven bursts from the M81 source were detected by CHIME during an on-source time of around 100 hours \citep{Kirsten2022}, which yields a burst rate of approximately 0.07 per hour above a fluence of 5\,Jy ms. \citet{Kremer2023} conclude that M87 is the best candidate for GC FRB detection. However, M87 is a strong radio source ($\approx 200$\,Jy at L-band) due to the supermassive black hole at its centre. We estimate this could reduce the sensitivity to FRBs by a factor of 4 or more (due to the increase in the system temperature) and certainly reduces the expected detection rate from such sources. A preferable choice is a radio-quiet galaxy with a large population of GCs. Therefore, we have selected M49 (RA: 12:29:46.80, Dec: +08:00:01.5), which is also known as NGC\,4472, a nearby radio-quiet ($\approx$ 0.2\,Jy) giant elliptical galaxy in the Virgo Cluster. It has a very rich retinue of GCs \citep[$\approx$ 7000, ][]{Harris2013} and lies at a similar distance to M87 \citep[$\approx$ 17 Mpc,][]{Blakeslee2009}. By using the same methodology as \citet{Kremer2023}, we estimate that a FAST multibeam survey of M49 of 50~hr has a $>$ 90\% chance of detecting a burst for $\alpha = -1.5$ \citep{Kremer2023} and 5 hr for $\alpha = -2.4$ \citep{Nimmo2023}, where $\alpha$ is the assumed power-law index of the FRB energies (i.e. slope of $log\mathrm{N}log\mathrm{S}$). 

\section{Observations} \label{sec:M49observation}
We conducted a series of observations of M49 as part of project PT2023\_0173. Three sessions were carried out with the FAST 19-beam receiver in SnapShotCal mode \citep{Han2021} on 8 October 2023, 15 December 2023, and 3 April 2024. Each observation had a total duration of 10,200\,s (\revt{$\approx$ 0.94}\,hr), of which the initial 120\,s were used for noise diode injection for calibration purposes. This yielded an on-source integration time of \revt{10,080}\,s per session, corresponding to a total integration of \revt{8.4}\,hr. The pulsar (psr) backend was used for the observation, with 19 beams employed. Observations were conducted with a centre frequency of 1.25\,GHz across 400 MHz of bandwidth, with all Stokes parameters recorded. The data has a sampling time of 49.152\,$\upmu$s and 4096 frequency channels (giving a frequency resolution of about $0.1$\,MHz).

\begin{figure}
\centering
	\includegraphics[width=\linewidth]{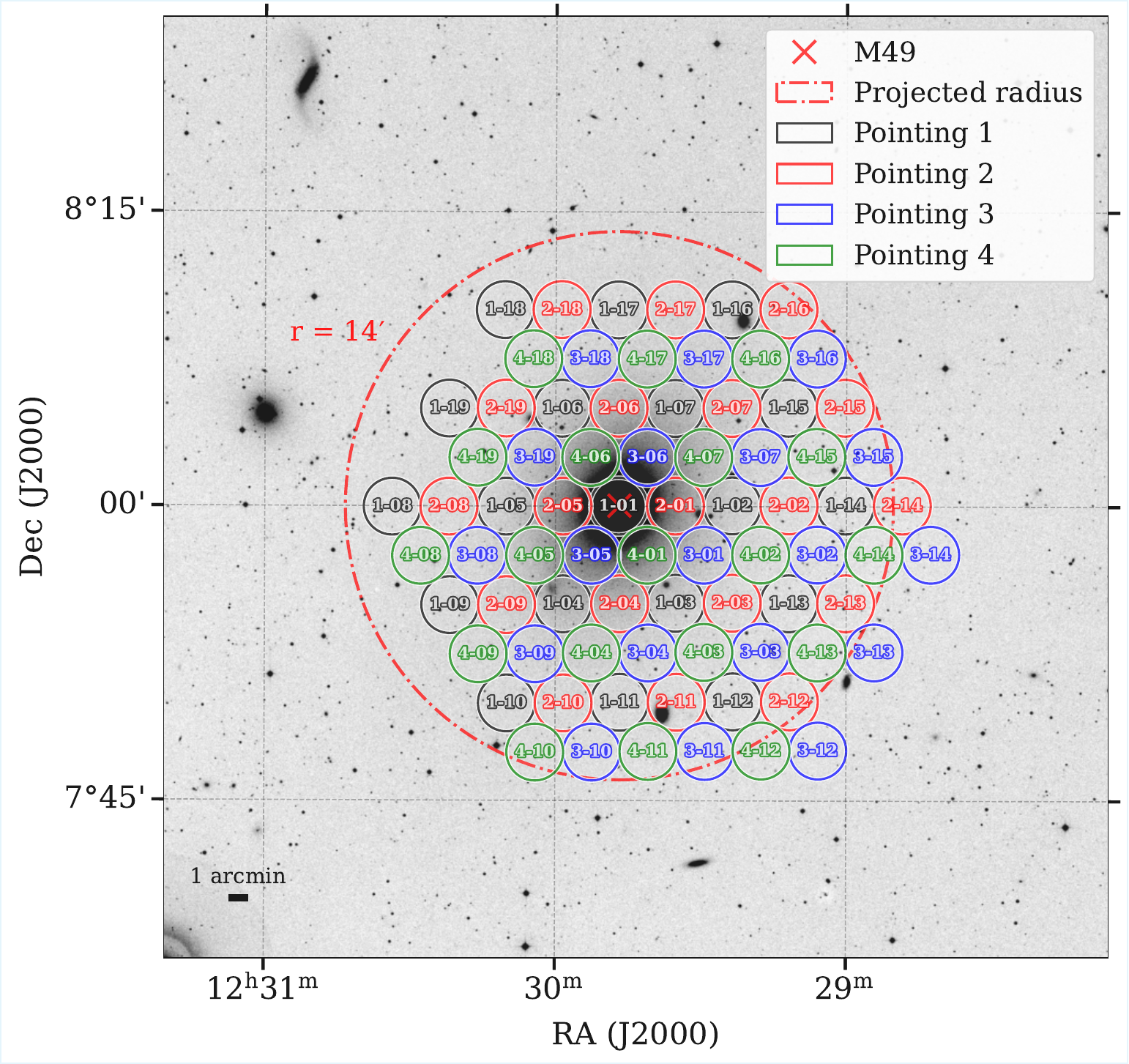}
    \caption{A $50\times50$\,arcmin DSS2 $r$-band image of M49. The red cross marks the centre of the galaxy, and the red dash-dotted circle indicates a radial distance of 14\,arcmin for reference. The \rev{half-power radius (1.45\,arcmin)} of the FAST 19-beam receiver is shown for pointings 1--4 in SnapShotCal mode, overplotted as black, red, blue, and green circles, respectively. Approximately 4230 GCs fall within the beam pattern.}
    \label{M49_snapshot_19beams}
\end{figure}

In Fig. \ref{M49_snapshot_19beams}, we show the sky coverage of the four pointings of the FAST 19-beam receiver in SnapShotCal mode \citep{Han2021}. The observation begins with pointing 1, centred at the central position of the galaxy, which is tracked for 5 minutes while data from all 19 beams are recorded. The beam pattern then shifts to pointing 2 and tracks for 5 minutes, and the same procedure is repeated for pointings 3 and 4. After completing pointing 4, the sequence returns to pointing 1, and the cycle continues until the full observation is finished. This means we have a phase-incoherent integration time of around 7480 s per beam on M49 for the full observation. \rev{Because the SnapShotCal pattern cycles through four pointings, a given GC is not observed continuously over the full 9\,hr observation, but only during the intervals when it lies within the beam pattern. We therefore define the effective exposure time per GC, $T_{\rm eff}$, as the total accumulated dwell time during which that GC is covered by the beams. For the present configuration, this corresponds to $T_{\rm eff}=\revt{2.1}$\,hr per GC over the full observation. The expected number of detected bursts can therefore be expressed as $\lambda = R\,N_{\rm GC}\,T_{\rm eff}$, where $R$ is the burst rate per GC, $N_{\rm GC}$ is the number of GCs within the beam footprint, and $T_{\rm eff}$ is the effective exposure per GC.} The four pointings cover a sky area of $\approx$ 0.1575 square degrees \citep{Han2021}. We plotted a dashed red circle with a projected radial distance of 14 arcmin around the galaxy, which has a similar coverage. To estimate the number of GCs that we can cover with this set-up, we use the surface density of the GCs in M49 that is shown in \citet{Hargis2014} where they fitted a S\'ersic function. They obtained a S\'ersic index of $3.6 \pm 0.6$. We perform an independent S\'ersic fit with \textsc{optimize.curve\_fit} function in \textsc{Scipy} \citep{scipy2020} and derive an index of $n = 3.8 \pm 0.6$ which is consistent with their result. Integrating our S\'ersic model from $r = 0$ to 14 arcmin, we estimate that the four pointings of FAST can enclose around $4230 \pm 65$ GCs throughout the observation. \rev{\rev{Under the simplifying assumption of uniform beam sensitivity across the covered region, each GC receives an effective exposure of $T_{\rm eff} = \revt{2.1}$\,hr in a geometric sense. In practice, the beam gain decreases away from the beam centre, so the true completeness is position dependent. This effect is accounted for in Section~6.5, where we incorporate beam sensitivity variations into the effective GC coverage and rate calculation.}}

\begin{figure}
\centering
	\includegraphics[width=\linewidth]{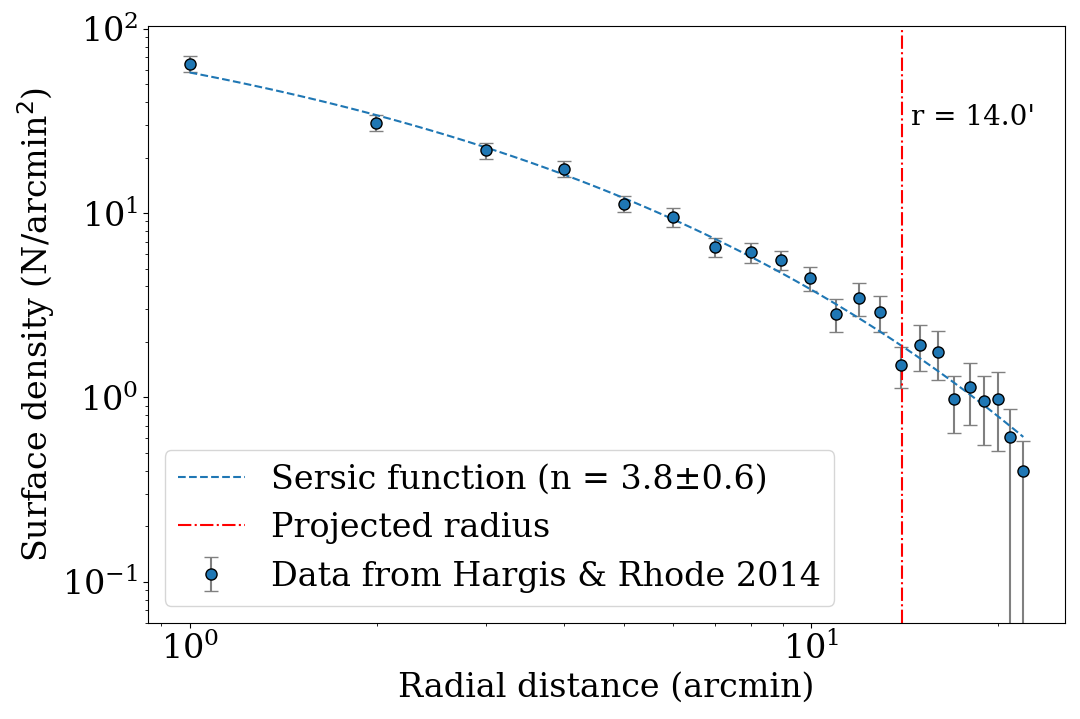}
    \caption{The surface density of GCs plotted against the radial distance for M49, shown on logarithmic axes. The data are from \citet{Hargis2014}. The fitted S\'ersic function and the radial distances of $r=14'$ from the galaxy centre are plotted with a blue dashed line and a red dash-dotted line.}
    \label{numberdensity_radial}
\end{figure}

\section{Data preprocessing and single-pulse search}
\label{sec:M49search}

The observational data consist of short FITS files with a duration of $\approx6.37$\,s each. To minimise the loss of signals at the boundaries between consecutive files, we concatenated 75 contiguous FITS files into a single 480\,s filterbank file. This resulted in twenty 480~s filterbanks and one shorter 408\,s filterbank for the last filterbank in a 3\,hr observation. To reduce computational cost and disk usage, all filterbanks were converted to 2-bit total-intensity (Stokes I) data with a time resolution of 98.304\,$\upmu$s.

We conducted a single-pulse search on the filterbanks using \textsc{TransientX} \citep{Men2024}, an efficient CPU-based pipeline employing sub-band dedispersion. The pipeline includes radio-frequency interference (RFI) mitigation, dedispersion, matched filtering, candidate clustering, and candidate visualisation. Pulses were searched over a dispersion measure (DM) range of 0--5000\,pc\,cm$^{-3}$ to accommodate that high DMs background transients may be detected. To suppress slow-varying RFI, we enabled the KadaneF filter and zero-DM matched filter using the option \texttt{-z kadaneF 8 4 zdot}. This filter mitigates temporally correlated RFI across frequency channels by identifying the time window with the maximum cumulative signal using a Kadane algorithm-based search. The two integers specify the pre-downsampling factors in time ($t_{\mathrm{d}}=8$) and frequency ($f_{\mathrm{d}}=4$), respectively. We also apply the zero-DM matched filter using the option \texttt{zdot}. This filter subtracts the correlation component between the data in each channel and the zero-DM profile. A dedispersion plan (DDplan) was generated using the \textsc{ddplan.py} script from the PulsaR Exploration and Search TOolkit \citep[\textsc{PRESTO};][]{Ransom2011}, based on the observational parameters (e.g. bandwidth, number of frequency channels, and time resolution). This approach trades a small amount of sensitivity loss for a substantial reduction in computational cost. \rev{The DDplan contains $N_{\mathrm{DM}}=16{,}646$ DM trials over the range $0$--$5000\,\mathrm{pc\,cm^{-3}}$. Since the DM spacing varies with DM in the DDplan, we do not use a single constant DM step size; instead, the step size is chosen to keep the S/N loss from residual intra-channel smearing below the adopted tolerance. The matched-filter search was performed using $N_{\mathrm{boxcar}}=32$ boxcar widths spanning from 0.1\,ms to 100\,ms. These values are used later in the trials-factor estimate in Section~\ref{sec:M49discussion}. The DDplan is also shown in Appendix~\ref{M49appendix} for visualisation.}

%, approximately logarithmically spaced, i.e. generated as a geometric progression in units of the sampling time with spacing set by the allowed S/N loss. 

%As an independent cross-check, we also performed a complementary search using the deep-learning-based RAdio Fast Transient Search pipeline \textsc{DRAFTS} \citep{Zhang2025}. This pipeline is optimised for FAST data and incorporates a CUDA-accelerated dedispersion algorithm, an object detection model to estimate the time of arrival (ToA) and DM of transient signals, and a binary classification model to reject false positives.

%DM estimation
%Expected rate from M49

\section{Results} \label{sec:M49results}

\rev{After processing a total of 29,940\,s (\revt{8.4}\,hr) of on-source data}, \textsc{TransientX} returned 23,415 and 1,348 candidates with signal-to-noise ratio $(\mathrm{S/N}) > 7.0$ and $> 10.0$, respectively, across the 19 beams. Candidates were visually inspected to identify potential astrophysical signals.

No significant detections with $\mathrm{S/N} > 10$ were identified in the search.  We therefore examined candidates with $7 < \mathrm{S/N} < 10$.  After excluding events clearly attributable to radio-frequency interference (RFI)  (examples are shown in Appendix~\ref{M49appendix2}), we retained 20 candidates exhibiting potentially astrophysical characteristics, including broadband emission and DMs exceeding the expected Milky Way contribution of $\mathrm{DM}_{\mathrm{MW}} \approx 30\,\mathrm{pc\,cm^{-3}}$ \citep{Cordes2002}.

These candidates were optimised using the \textsc{replot\_fil} in \textsc{TransientX}, which applies finer DM grids and refined processing \citep{Men2024}. Following this procedure, one candidate originally detected at $\mathrm{S/N} = 7.4$ increased to $\mathrm{S/N} = 8.6$, making it the most significant event identified in the full \revt{8.4}\,hr data set.

This candidate has a fluence of $\rev{14.5}\,\mathrm{mJy\,ms}$, a pulse width of $0.7\,\mathrm{ms}$, and a DM of $412.2\,\mathrm{pc\,cm^{-3}}$. It was detected in beam~10 ($\sim 15'$ offset from the galaxy centre), with no simultaneous detections in adjacent beams. No RFI storm was observed within 10\,s of the event. Although elevated RFI activity occurred approximately 40\,s earlier which produced a temporary increase in candidate counts by nearly two orders of magnitude, this activity is unlikely to be directly associated with the event.

\begin{figure*}
\centering
\includegraphics[width=0.464\linewidth]{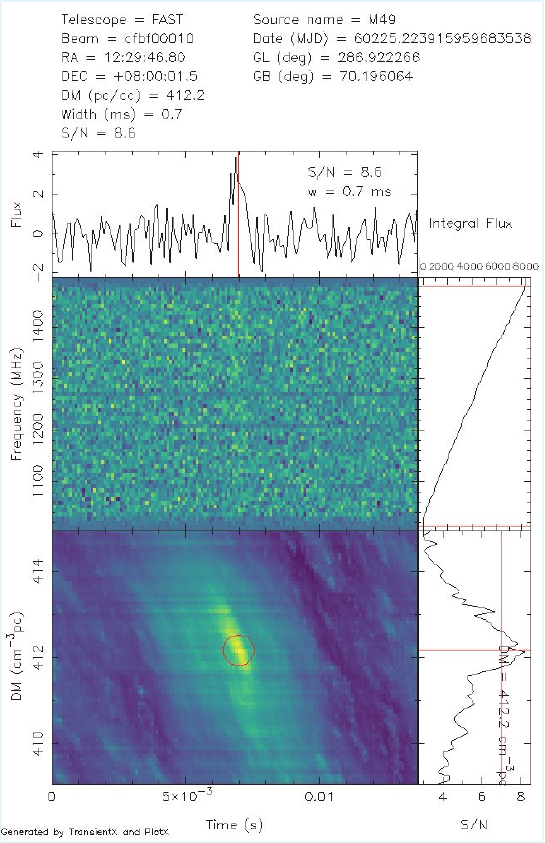}
\includegraphics[width=0.48\linewidth]{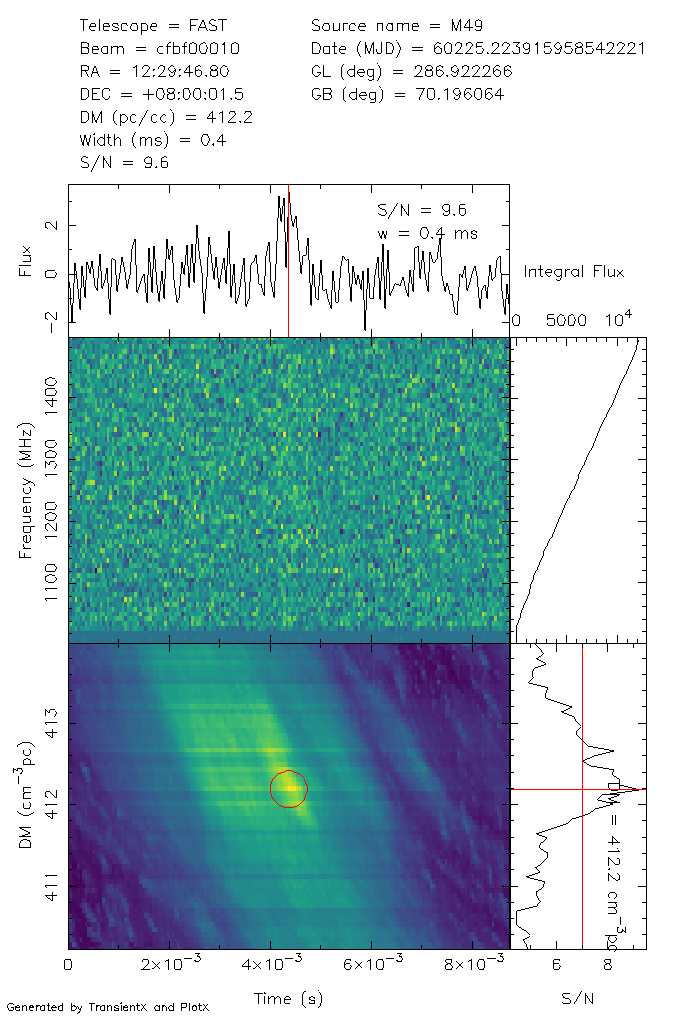}
\caption{Most significant single-pulse candidate identified in the \revt{8.4}\,hr FAST observation of M49. Left: Detection in the search-resolution data (2-bit, 98.304\,$\upmu$s). Right: Reprocessing at full resolution (8-bit, 49.152\,$\upmu$s, full polarisation). The panels show (from top to bottom): integrated pulse profile, dynamic spectrum, $\mathrm{S/N}$ distribution in the DM–time plane, accumulated flux, and $\mathrm{S/N}$ as a function of DM. The full-resolution reprocessing yields a slightly higher peak $\mathrm{S/N}$, which may be due to reduced temporal smearing and quantisation loss.}
\label{fig:M49_cand}
\end{figure*}

We further processed the candidate at full resolution (8-bit sampling, $49.152\,\upmu$s, full polarisation), yielding a peak $\mathrm{S/N}$ of 9.6. The modest increase in significance is likely attributable to reduced temporal smearing and the removal of quantisation loss. Independent verification using \textsc{pdmp} in \textsc{PSRCHIVE} \citep{Hotan2004} yields peak $\mathrm{S/N}$ values of 7.8 in the time-scrunched data and 8.3 at full time resolution. The \textsc{pdmp}-derived $\mathrm{S/N}$ values differ by $10-15\%$ from the \textsc{replot\_fil} estimates, consistent with expected variations arising from differences in $\mathrm{S/N}$ definition. The search-resolution and full-resolution plots of the candidate are shown in Figure~\ref{fig:M49_cand}. A detailed assessment of its statistical significance and possible astrophysical interpretations is presented in Section~\ref{sec:M49discussion}.

\section{Discussion} \label{sec:M49discussion}

\subsection{Noise statistics and false-alarm probability}
Fluctuations in background noise (e.g.\ Gaussian thermal noise) often limit the detectability of FRBs and pulsars. For a signal to be considered significant, its $\mathrm{S/N}$ must exceed the largest noise fluctuations expected over the full search. For Gaussian thermal noise, the probability of obtaining a fluctuation above a threshold $c_{\mathrm{thresh}}$ is given by 

\begin{equation}
p_{\mathrm{FA}}(c_{\mathrm{thresh}})
=
\int_{c_{\mathrm{thresh}}}^{\infty}
\frac{1}{\sqrt{2\pi}}
\exp\!\left(-\frac{c^{2}}{2}\right)\,
\mathrm{d}c
=
\frac{1}{2}
\operatorname{erfc}
\left(
\frac{c_{\mathrm{thresh}}}{\sqrt{2}}
\right),
\label{eq:noise_tail}
\end{equation}
\citep{Condon2016}

For $c_{\mathrm{thresh}} = 7.4$, the single-trial false-alarm probability is

\begin{equation}
p_{\mathrm{FA}}(7.4)
=
\frac{1}{2}
\operatorname{erfc}
\left(
\frac{7.4}{\sqrt{2}}
\right)
=
6.8 \times 10^{-14}.
\end{equation}

The total number of independent trials in the search can be expressed as
\begin{equation}
N_{\mathrm{trials}} =
N_{\mathrm{samples}} \times
N_{\mathrm{DM}} \times
N_{\mathrm{boxcar}} \times
N_{\mathrm{beam}},
\end{equation}
where $N_{\mathrm{samples}}$ is the number of independent time samples, $N_{\mathrm{DM}}$ is the number of DM trials, $N_{\mathrm{boxcar}}$ is the number of boxcar widths used in the search, and $N_{\mathrm{beam}}$ is the number of beams searched.

For this 29,940\,s observation, we have
$N_{\mathrm{samples}} = 29,940\,s /98.304\,\upmu s = 3.05 \times 10^{8}$,
$N_{\mathrm{DM}} = 16,646$ (see Appendix \ref{M49appendix}),
$N_{\mathrm{boxcar}} = 32$, and
$N_{\mathrm{beam}} = 19$,
resulting in a total of
\begin{equation}
N_{\mathrm{trials}} = 2.90 \times 10^{15}.
\end{equation}
The expected number of false positives at an $\mathrm{S/N}$ threshold of 7.4 is therefore
\begin{equation}
N_{\mathrm{false}} =
N_{\mathrm{trials}} \times
p_{\mathrm{FA}}(7.4)
=
2.90 \times 10^{15}
\times
6.8 \times 10^{-14}
\approx 197.
\end{equation}

Note that this estimate assumes statistically independent trials, \revt{whereas in practice there are correlations between neighbouring DM trials, overlapping boxcar filters, and adjacent time samples.} The corresponding false-alarm estimate should therefore be regarded as approximate; treating all trials as independent makes the quoted value conservative.

The large expected false-positive rate indicates that such triggers are anticipated from thermal noise once the trials factor is taken into account. We therefore conclude that the most significant trigger in the dataset is statistically consistent with Gaussian noise. Nevertheless, we retain this event for further discussion below in order to assess its properties and to explore the implications if it were astrophysical in origin.

\subsection{DM estimation for an M49 GC}
\label{sec:DM}

\begin{figure}
\centering
	\includegraphics[width=\linewidth]{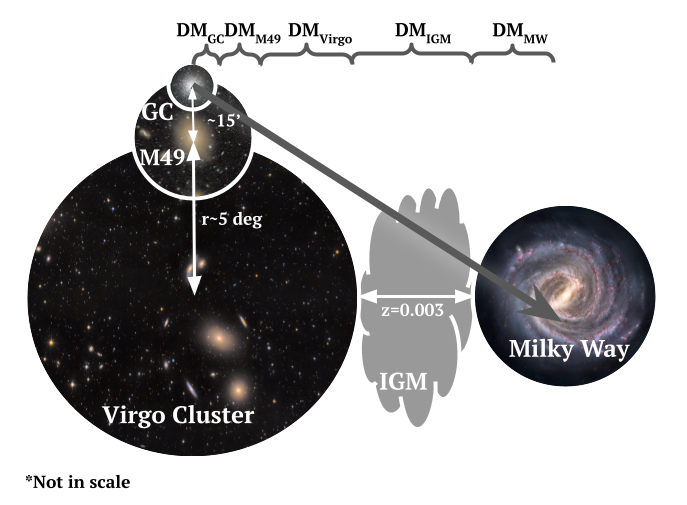}
    \caption{
    Schematic illustration of the DM budget for an FRB originating from a GC associated with M49, at a projected offset of $\approx15'$ ($\approx73$~kpc) from the galaxy centre. The grey arrow is the line-of-sight from the GC to the observer. The observed DM is decomposed into contributions from the Milky Way disk and halo ($\mathrm{DM}_{\mathrm{MW}}$), the Virgo intracluster medium in the outskirts of the Virgo B subcluster ($\mathrm{DM}_{\mathrm{Virgo}}$), the hot X-ray halo of M49 modelled by a $\beta$-profile ($\mathrm{DM}_{\mathrm{M49}}$) and the GC/local environment ($\mathrm{DM}_{\mathrm{GC}}$). Background images adapted from: (i) Milky Way by Nick Risinger (\url{https://en.wikipedia.org/wiki/File:Milky_Way_Galaxy.jpg}); (ii) Virgo Cluster / M49 by RubinObs/NOIRLab/SLAC/NSF/DOE/AURA (\url{https://noirlab.edu/public/images/noirlab2521a/}); (iii) GC by NASA/ESA/STScI/A. Sarajedini (\url{https://science.nasa.gov/mission/hubble/science/explore-the-night-sky/hubble-messier-catalog/messier-2/}).
}
    \label{fig:DM_plot}
\end{figure}

The total DM for a radio signal originating from a GC associated with M49 can be expressed as
\begin{equation}
\mathrm{DM}_{\mathrm{total}} =
\mathrm{DM}_{\mathrm{MW}} +
\mathrm{DM}_{\mathrm{IGM}} +
\mathrm{DM}_{\mathrm{Virgo}} +
\mathrm{DM}_{\mathrm{M49}} +
\mathrm{DM}_{\mathrm{GC}},
\label{eq:dm_M49}
\end{equation}
where $\mathrm{DM}_{\mathrm{MW}}$, $\mathrm{DM}_{\mathrm{IGM}}$, $\mathrm{DM}_{\mathrm{Virgo}}$, $\mathrm{DM}_{\mathrm{M49}}$, and $\mathrm{DM}_{\mathrm{GC}}$ represent contributions from the Milky Way (disk + halo), the intergalactic medium (IGM), the Virgo intracluster medium (ICM), the hot halo of M49, and the GC/local environment, respectively (see Figure \ref{fig:DM_plot}).

The expected Milky Way contribution along the line of sight to M49 is approximately $29~\mathrm{pc\,cm^{-3}}$\rev{, $27~\mathrm{pc\,cm^{-3}}$} and $21~\mathrm{pc\,cm^{-3}}$ based on the NE2001 \citep{Cordes2002}\rev{, NE2025 \citep{Ocker2026}} and YMW16 \citep{Yao2017} electron-density models, respectively. The Milky Way halo contributes an additional $\approx 25$--$80~\mathrm{pc\,cm^{-3}}$ \citep{Prochaska2018, Shull2018, Prochaska2019, Yamasaki2020}. We therefore adopt $\mathrm{DM}_{\mathrm{MW}} = 46$--$109~\mathrm{pc\,cm^{-3}}$.

M49 is a nearby galaxy with a low redshift $z \approx 0.003$, such that the IGM contribution is small. Using the Macquart relation \citep{Macquart2020}, we estimate $\mathrm{DM}_{\mathrm{IGM}} \approx 3~\mathrm{pc\,cm^{-3}}$, and therefore treat it as negligible.

M49 resides in the Virgo B subcluster, located in the outskirts of the Virgo cluster. Adopting representative electron densities for the outer ICM of Virgo \citep[e.g.][]{Nulsen1995}, we estimate an intracluster DM contribution of $\mathrm{DM}_{\mathrm{Virgo}} \approx 30$--$50~\mathrm{pc\,cm^{-3}}$. As a consistency check, applying Eq.~(5) of \citet{Planck2016} yields a comparable value of $\mathrm{DM}_{\mathrm{Virgo}} \approx 50~\mathrm{pc\,cm^{-3}}$. We therefore adopt $\mathrm{DM}_{\mathrm{Virgo}} = 30$--$50~\mathrm{pc\,cm^{-3}}$, noting that this range accounts for the uncertainty in the burst location along the line of sight through the Virgo cluster (i.e. whether the source lies on the near or far side of the ICM).

M49 is a giant elliptical galaxy and therefore lacks a star-forming disk. Although it contains some central dust and cold gas structures, these are confined to the inner 
$\sim$kpc region \citep{Thomas1986, Kraft2011}. Consequently, the host-galaxy DM contribution, 
$\mathrm{DM}_{\mathrm{M49}}$, is expected to be dominated by the diffuse hot halo. \citet{Su2019} reported a hot X-ray--emitting gas halo around M49 and modelled its electron density profile using a $\beta$-model:

\begin{equation}
n_{\mathrm{e}}(r) = n_{0}
\left[ 1 + \left( \frac{r}{r_{\mathrm{c}}} \right)^{2} \right]^{-3\beta/2},
\label{eq:beta_model}
\end{equation}
where $n_{0}$, $r_{\mathrm{c}}$, and $\beta$ are the central electron density, core radius, and slope parameter, respectively.

Integrating this density profile along the line of sight from a typical GC location ($r_s \approx 20$--$60~\mathrm{kpc}$) to the outer boundary of the halo yields 
$\mathrm{DM}_{\mathrm{M49}} \approx 10$--$40~\mathrm{pc\,cm^{-3}}$.
The candidate reported in Section~\ref{sec:M49results} was detected in beam~10, corresponding to a projected offset of $\approx 15$~arcmin ($\approx 73$~kpc) from the centre of M49. At this larger impact parameter, the hot-halo contribution is expected to be smaller; we adopt $\mathrm{DM}_{\mathrm{M49}} \approx 5$--$20~\mathrm{pc\,cm^{-3}}$. This range accounts for the uncertainty in the burst location along the line of sight through M49.

For comparison, if an FRB originated near the dense central region of M49, the much higher electron densities predicted by the $\beta$-model would yield a substantially larger host contribution. Integrating the profile from the core to the halo boundary implies 
$\mathrm{DM}_{\mathrm{M49,core}} \sim 50-150~\mathrm{pc\,cm^{-3}}$, 
with the exact value depending on the adopted $\beta$-model parameters and halo extent.

Finally, GCs are gas-poor systems \citep{Roberts1960, Knapp1996}, such that any additional DM contribution internal to the GC is expected to be negligible compared to the diffuse halo/ICM terms. Thus, we treat $\mathrm{DM}_{\mathrm{GC}}$ as negligible. Summing the individual components, we estimate
\begin{equation}
\mathrm{DM}_{\mathrm{total}} \approx 81\text{--}179~\mathrm{pc\,cm^{-3}}.
\end{equation}
The measured DM of the candidate is $412.2~\mathrm{pc\,cm^{-3}}$, which significantly exceeds the expected DM budget for an M49 GC origin. This suggests that the candidate is unlikely to be associated with an M49 GC. However, the DM budget relies on simplified assumptions for the host-galaxy halo and intracluster medium, and the associated uncertainties prevent us from completely ruling out a GC origin. We also note that an additional DM contribution from the immediate vicinity of the progenitor (e.g. a nebula or magnetar wind) is uncertain and not included explicitly.

\subsection{Localisation uncertainty}
Localisation precision is crucial for determining the exact location of the FRB within its host galaxy \citep{Caleb2017, Nimmo2022, Hewitt2024} to determine its origin/progenitor. For FRB\,20200120E, \citet{Kirsten2022} achieved a localisation uncertainty of 0.7 and 0.4 mas in right ascension and declination, respectively. Since FAST is not an interferometer, the best localisation we can constrain is within the \rev{half-power radius} of a particular beam ($r = \rev{1.45}$\,arcmin), which corresponds to an uncertainty of approximately \rev{2.9} arcmin in both coordinates. \citet{Jordan2005} estimated the mean half-light radius of M49 GCs as $r_{\text{h}} = 0.033 \pm 0.001$ arcsec. This is nearly five orders of magnitude smaller than the FAST beam. Thus, even if an FRB originates within a GC of M49, FAST cannot pinpoint it to an individual cluster. However, the surface density of M49 GC is high (ranges from around 5 to more than 60 GCs/arcmin$^2$), meaning that a single FAST beam will often contain numerous GCs. Therefore, it remains possible that any detection in this observation is associated with a GC, even though a unique GC host cannot be identified. To confirm the association definitively, interferometric follow-up with sub-arcsecond or milliarcsecond precision would be required.

\subsection{Detection rate of a background FRB}

Since our localisation precision is insufficient to associate the detection with a specific GC, and the measured DM exceeds the expected DM budget for M49, there remains the possibility that the event is instead a background FRB \citep[e.g.,][]{Carli2024}. Here, we examine this possibility quantitatively.

The Commensal Radio Astronomy FAST Survey (CRAFTS) has discovered a total of 10 FRBs over approximately 3000\,hr of observing time \citep{Zhu2020, Niu2021, Niu2022}. This corresponds to an average detection rate of $\sim$1 FRB per 300\,hr, or equivalently $\sim 0.0033$~FRB\,hr$^{-1}$. Adopting this rate, the expected number of background FRBs detectable during our \rev{29,940\,s (\revt{8.4}\,hr) on-source} observation is

\begin{equation}
N_{\mathrm{exp}} \approx 0.0033 \times \revt{8.4} \approx \revt{0.028} .
\end{equation}

This implies that while the probability of detecting a background FRB within our observing time is low, it is not negligible, and thus the candidate event could plausibly originate from an unrelated background source. \revt{This estimate should be regarded as an order-of-magnitude approximation only, as differences in observing strategy, sensitivity, sky coverage, and pipeline completeness between surveys limit the precision of a direct comparison with the CRAFTS detection rate.}

\subsection{Search sensitivity and upper limit} \label{sec:M49upperlimit}
The $8.6\sigma$ event is an intriguing candidate; however, its astrophysical nature cannot be robustly established. We therefore conservatively assume that no significant FRB or FRB-like pulses were detected and derive the sensitivity limit of our search using the radiometer equation.
\begin{equation}
S_{\min,\mathrm{pulse}}
=
\frac{\left(\mathrm{\mathrm{S/N}}\right)\,\left(T_{\mathrm{sys}}\right)\,\beta}
{G\,\sqrt{n_{\mathrm{p}}\,\Delta\nu\,w_{\mathrm{broadened}}}}
\end{equation}
\citep[e.g.,][]{McLaughlin2003, Carli2024}, where $\mathrm{S/N}$ is the $\mathrm{S/N}$ threshold, $T_{\mathrm{sys}}$ is the system temperature including the sky temperature, $\beta$ is the digitisation correction factor \citep{Kouwenhoven2001}, $G$ is the gain, $n_{\rm p}$ is the number of summed polarisations, $\Delta\nu$ is the effective observing bandwidth, and $w_{\mathrm{broadened}}$ is the observed pulse width. As part of our calculation, we also account for sampling time and DM smearing \citep{Lorimer2004}. Therefore,

\begin{equation}
w_{\mathrm{broadened}}
=
\sqrt{t_\mathrm{int}^2 + t_\mathrm{s}^2 + \tau_\mathrm{DM}^2\rev{ + t_\mathrm{scat}^2}}
\end{equation}

We assume an intrinsic pulse width, $t_\mathrm{int}$, of $1$\,ms at $\mathrm{DM}=200\,\mathrm{pc\,cm^{-3}}$. For our search, the time resolution $t_\mathrm{s}$ is $98.304\,\upmu$s and the dispersive smearing $\tau_\mathrm{DM}$ across individual frequency channels is 0.083\,ms. \rev{We neglect scattering in this fiducial estimate and set $t_\mathrm{scat}=0$. This is reasonable for a reference sensitivity estimate, as strong scattering is not expected for all GC environments and the known GC-associated source FRB\,20200120E exhibits many narrow bursts at L-band \citep{Kirsten2022, Nimmo2023}. With this assumption, $w_{\mathrm{broadened}} \approx 1.01$\,ms.}

\rev{We adopt a detection threshold of $(\mathrm{S/N})=10$, $T_\mathrm{sys}=21$\,K, $n_{\rm p}=2$ for Stokes I, $\beta=1.06$ for 2-bit data \citep{Kouwenhoven2001}, and an effective bandwidth of $\Delta\nu=400$\,MHz.}

\rev{To account for the variation in sensitivity across the 19 FAST beams, we scale the gain of each beam relative to a reference value of $G=16.12 \pm 0.25$\,K\,Jy$^{-1}$ using the measured gain ratios at a central frequency of 1250 MHz (see Table~5 of  \citealt{Jiang2020}). The gain of beam $i$ is therefore given by $G_i = r_i \times 16.12$\,K\,Jy$^{-1}$, where $r_i$ is the tabulated ratio relative to the central beam.}

\rev{The single-pulse peak flux density sensitivity for beam $i$ is then
\begin{equation}
S_{{\rm pulse,peak},i}
=
\frac{(\mathrm{S/N})\,T_{\rm sys}\,\beta}
{G_i \sqrt{n_{\rm p}\,\Delta\nu\,w_{\rm broadened}}}.
\end{equation}}

\rev{For the central beam ($r_i=1$), this yields
\begin{equation}
S_{{\rm pulse,peak},01}
\approx
\frac{10 \times 21\,\mathrm{K} \times 1.06}
{16.12\,\mathrm{K/Jy} \sqrt{2 \times 400\,\mathrm{MHz} \times 1.01\,\mathrm{ms}}}
\approx 15.5\,\mathrm{mJy}.
\end{equation}}

\rev{Across the 19 beams, the gain ratios at 1250 MHz range from $r_i \approx 0.86$ to $1.00$ (Table~5 of  \citealt{Jiang2020}), corresponding to peak flux density sensitivities of $\sim 15.5$--$18.0$\,mJy. Averaging over all beams gives an effective (beam-averaged) sensitivity of $\sim 16.5$\,mJy for a 1.01 ms burst.}

\begin{figure}
\centering
	\includegraphics[width=\linewidth]{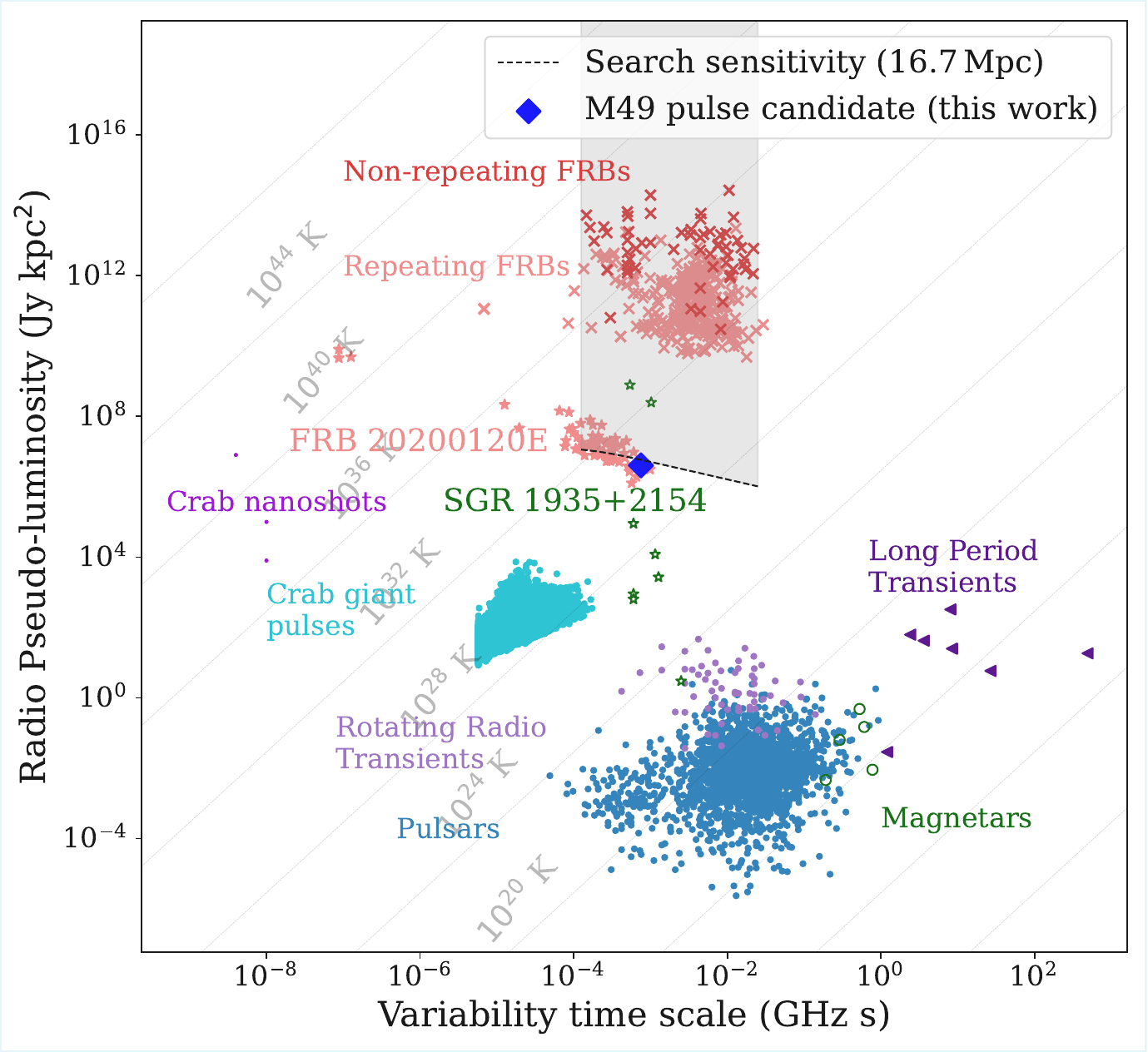}
    \caption{Radio transient phase space diagram created using the \texttt{TransientPhaseSpace} GitHub repository (\url{https://github.com/lgspitler/TransientPhaseSpace}) with data from \citet{Nimmo2022a}, \rev{\citet{Kirsten2022} and \citet{Nimmo2023}}, showing the distribution of various transient classes in parameter space. The candidate identified in this work is overplotted \rev{(assuming a distance of 16.7\,Mpc)} and lies close to the location of FRB\,20200120E in this parameter space. \rev{The dashed line indicates the survey sensitivity as a function of burst width at a distance of 16.7\,Mpc.}}
    \label{fig:Lum_time}
\end{figure}

This represents the limiting peak flux density for millisecond-duration bursts detectable in our \rev{29{,}940\,s (\revt{8.4}\,hr)} FAST observation. This corresponds to a fluence sensitivity lower limit of $F_{\rm min} \approx S_{\rm pulse,peak} w_{\mathrm{broadened}} \approx \rev{16.5}\,\mathrm{mJy\,ms}$. \rev{Since the fluence threshold scales approximately as $\sqrt{W}$, where $W \equiv w_{\mathrm{broadened}}$ is the observed pulse width, wider bursts would require correspondingly higher fluences for detection. For example, a 10\,ms burst would require a fluence of $\sim 16.5\sqrt{10}\approx 52\,\mathrm{mJy\,ms}$, compared to $\approx 16.5\,\mathrm{mJy\,ms}$ for a 1\,ms burst.} We place our search sensitivity with the same average sensitivity (16.5\,mJy\,ms) as a function of intrinsic burst width onto a radio transient phase space diagram in Figure~\ref{fig:Lum_time} using a distance of 16.7\,Mpc. \rev{We also plot the repeating bursts of FRB\,20200120E from \citet{Kirsten2022} and \citet{Nimmo2023}, together with the nanosecond-scale sub-structures reported by \citet{Nimmo2022a}, which are shown for comparison but do not represent independent burst events.} The candidate we found in Section \ref{sec:M49results} is also plotted, comparing it with various classes of known radio transients. Interestingly, the candidate lies close to the location of FRB\,20200120E, the FRB associated with a GC in M81. 

\rev{A key question is whether bursts from FRB\,20200120E would be detectable at the distance of M49. FRB~20200120E, located in a GC in M81, exhibits many bursts with fluences well below $1~\mathrm{Jy\,ms}$. Scaling these bursts from the distance of M81 ($\sim 3.6$~Mpc) to that of M49 ($\sim 16.7$~Mpc) reduces their observed fluences by a factor of $(3.6/16.7)^2 \approx 0.047$. Using our beam-averaged FAST fluence threshold of $16.5~\mathrm{mJy\,ms}$ for a $1~\mathrm{ms}$ burst, we find that the majority ($56/63$) of published FRB~20200120E bursts \citep{Kirsten2022, Nimmo2023} would fall below our detection limit. However, many bursts from FRB~20200120E are significantly narrower than $1~\mathrm{ms}$. Considering the subset of $58$ bursts with measured widths and accounting for the $\sqrt{W}$ dependence of the fluence threshold, we find that $31$ out of $58$ bursts would remain detectable, while a substantial fraction ($27/58$) would still fall below threshold \citep{Kirsten2022, Nimmo2023}. Our search is therefore primarily sensitive to the brighter portion of an FRB~20200120E-like burst distribution, and does not strongly constrain the presence of a lower-luminosity repeating source population in M49 GCs.}

\rev{To account for the variation in sensitivity across the beams, we introduce an effective number of GCs. 
Assuming a Gaussian beam response $B(x) = A \exp(-x^2)$, the half-power radius scales as $x_{\rm HP} \propto \sqrt{\ln(A/0.5)}$, where $A$ is the relative peak sensitivity. Adopting a mean sensitivity of $A=0.94$ due to gain variation \citep{Jiang2020} reduces the effective area by a factor of $\ln(0.94/0.5)/\ln 2 \approx 0.91$. 
This corresponds to an effective number of GCs of $N_{\rm GC,eff} \approx 0.91 \times 4230 \approx 3850$.} \revt{This correction should be regarded as an approximate beam-averaged completeness correction, as it assumes an idealised Gaussian beam response and uses the mean relative beam sensitivity to estimate the reduction in effective survey area.}
Since no unambiguous FRB-like events were detected in our search ($N=0$), we derive an upper limit on the event rate using \rev{Poisson statistics}. For zero observed events, the \rev{one-sided $2\upsigma$ (95\%) upper limit corresponds to an expected number of events of $N_{\rm UL}=3.78$ \citep[Table~1 of][]{Gehrels1986}. Given an effective exposure of $T_{\rm eff} \approx \revt{2.1}$\,hr per GC and an effective number of $\approx 3850$ GCs}, this yields an upper limit on the burst rate of
\begin{equation}
\rev{R_{\rm UL} \approx \frac{N_{\rm UL}}{T_{\rm eff} N_{\rm GC,eff}}
\approxeq \frac{3.78}{(\revt{2.1~{\rm hr}})\times 3850}
\approx \revt{4.7}\times 10^{-4}~{\rm FRB~GC^{-1}~hr^{-1}},}
\end{equation}
i.e., of order $\approx 10^{-4}$ bursts per GC per hour at our search sensitivity. \rev{We note that this derivation implicitly assumes a detection efficiency of unity ($\epsilon \approx 1$). In practice, the completeness depends on several factors, including the burst width distribution, dispersive smearing, scattering, beam sensitivity variation, and the efficiency of the search pipeline.} 

\rev{The variation in beam sensitivity across the FAST multibeam footprint is partially accounted for through the use of beam-dependent gain factors in our sensitivity estimate. The remaining effects are not explicitly modelled in this analysis. In particular, the quoted sensitivity is referenced to a fiducial burst width of 1\,ms, and broader bursts would require higher fluences for detection due to the $\sqrt{W}$ scaling. The efficiency of our search pipeline is not formally quantified through injection-and-recovery simulations. However, \citet{Men2024} have performed semi-analytical framework, which proves that the S/N degradation caused by pipeline parameter mismatches is strictly bounded and relatively minor. Also, \textsc{TransientX} pipeline \citep{Men2024} has been widely used in FRB searches \citep[e.g.,][]{Carli2024, Eppel2025, Shaji2026}. We therefore expect high detection efficiency for bright bursts above our threshold. The derived rate limit should therefore be regarded as an optimistic constraint applicable only to detectable bursts above our fluence threshold of $\approx 16.5\,\mathrm{mJy\,ms}$.}

For comparison, the repeating GC-associated source FRB~20200120E in M81 is known to undergo episodes of enhanced activity with multiple bursts detected within hour-long timescales \citep[e.g.][]{Bhardwaj2021,Kirsten2022,Nimmo2023}. \rev{Our non-detection therefore indicates only that no bright bursts were detected during our observing window. This does not rule out fainter repeaters, episodic sources, or sources with low duty cycles which bursts would fall below our detection threshold or occur outside the observing window.}

The expected detection rate depends on the assumed burst luminosity-function slope, which was estimated in Section~\ref{sec:M49sampleselection} using the burst sample of FRB\,20200120E in M81. We estimated that for a slope $\alpha=-1.5$ approximately $50$~hr of on-source integration would be required to expect one detectable burst, whereas for $\alpha=-2.4$ the required exposure would decrease to $\sim 5$~hr. However, that estimate is derived from a single repeating source observed during episodes of enhanced activity and is therefore subject to small-number biases. Consequently, any constraint on $\alpha$ inferred from our non-detection is necessarily model-dependent.

If we assume that the required exposure time scales approximately exponentially with $\alpha$, consistent with the two anchor points above, we may adopt the empirical relation
\begin{equation}
T(\alpha) = 50 \times 10^{\,\alpha - (-1.5)} \,\text{hr}.
\end{equation}
Requiring that $T(\alpha)$ equals our total \rev{on-source} time
($T=\revt{8.4}$~hr) gives
\begin{equation}
\alpha = \log_{10}(\revt{8.4}/50) - 1.5 \approx \rev{-2.3}.
\end{equation}
Under this illustrative scaling, slopes significantly steeper than $\alpha\sim \rev{-2.3}$ would predict $\gtrsim 1$ detectable burst within \revt{8.4}~hr and are therefore mildly disfavoured.

\rev{We emphasise that this estimate is purely illustrative and highly model-dependent. It is based on a single repeating source, assumes a specific empirical scaling with $\alpha$, and does not account for duty-cycle effects or population diversity.} Our non-detection does not provide a strong constraint on the intrinsic luminosity-function slope, but suggests that extremely steep slopes are unlikely if the M81 source is representative of the GC population. More robust constraints will require observations of a larger sample of GC-associated FRBs. Overall, our rate limit implies that if FRB sources similar to FRB\,20200120E exist in M49, they must either be intrinsically rare within the GC population or active only during a limited duty cycle within our observing window.

\section{Conclusion} \label{sec:M49conclusion}

We conducted a \rev{9-h FAST observation (\revt{8.4} hr on source)} to search for FRBs and FRB-like single pulses from the GCs of the giant elliptical galaxy M49, covering approximately $4230 \pm 65$ GCs \rev{($T_{\rm eff}=\revt{2.1}$~hr per GC)}. This represents one of the deepest single-pulse searches to date targeting the largest GC population in a single nearby galaxy. A comprehensive single-pulse search was performed using \textsc{TransientX} over a DM range of $0$--$5000\,\mathrm{pc\,cm^{-3}}$. Our main results can be summarised as follows:
\begin{enumerate}
\item No unambiguous astrophysical FRBs were detected. The most significant trigger identified in the dataset reached a post-processed $\mathrm{S/N}$ of 8.6, with a measured fluence of $\rev{14.5}\,\mathrm{mJy\,ms}$, a width of $0.7~\mathrm{ms}$, and a DM of $412.2\,\mathrm{pc\,cm^{-3}}$. However, once the thermal noise false-positive rate and trials factor are taken into account, such triggers are expected from Gaussian noise, and the event is therefore not considered a statistically significant astrophysical detection.

\item Interpreted hypothetically as an astrophysical burst, its DM exceeds the expected contribution from M49 (upper limit $\approx 179\,\mathrm{pc\,cm^{-3}}$), suggesting that it is unlikely to be associated with a M49 GC. While the probability of detecting a background FRB within the observing time and FoV is low ($\sim 0.03$ FRBs in \revt{8.4}~hr), it is non-negligible.

\item Assuming no confirmed single-pulse detections, we derive the sensitivity limit of our search using the radiometer equation. For an intrinsic pulse width of 1\,ms at $\mathrm{DM}=200\,\mathrm{pc\,cm^{-3}}$, we obtain a single-pulse peak flux-density sensitivity of \rev{$\approx 16.5$\,mJy, corresponding to a fluence threshold of $\approx 16.5$\,mJy\,ms, and a one-sided 95\% upper limit of $4.4\times 10^{-4}~\mathrm{FRB~GC^{-1}~hr^{-1}}$ for bursts above this threshold}.
\end{enumerate}

Although no FRBs were conclusively detected, these observations place meaningful upper limits on bright FRB emission from GCs owing to the exceptional sensitivity of FAST. \rev{No bright bursts above our detection threshold were detected during the observing window. These constraints are therefore sensitivity-limited and apply only to bursts above our detection threshold.} In contrast to the young, star-forming hosts of many FRBs, the absence of detectable bright bursts from the extensive GC population of M49 is consistent with scenarios in which the dominant FRB formation channel is associated with young magnetars, while still allowing for a rarer or more weakly bursting GC-associated channel such as FRB\,20200120E. \rev{This places only weak constraints on the intrinsic FRB occurrence rate in GCs, and remains consistent with scenarios involving episodic activity, low duty cycles, or populations dominated by bursts below our detection threshold.} If FRB\,20200120E represents a GC-associated channel, it may correspond to a rare or short-lived evolutionary phase rather than a common outcome of GC evolution.

At the same time, given the FRB occurrence rate estimated in Sections~\ref{sec:M49sampleselection} and \ref{sec:M49upperlimit}, the absence of detections in an \revt{8.4\,hr} observation is not unexpected. For a luminosity-function slope of $\alpha=-1.5$, approximately 50~hr of on-source integration would be required to expect a single detectable event, making a non-detection in \revt{8.4\,hr} entirely consistent with this model. While extremely steep slopes would predict a higher detection probability, our data do not provide a strong constraint on the intrinsic luminosity-function parameters. 

Longer-term monitoring with FAST or other wide-field instruments \citep{CHIME2018, Lin2022, Ho2023, Staveley-Smith2026} capable of covering large GC populations will therefore be essential for placing stronger constraints on FRB occurrence rates in GCs and nearby extragalactic environments. An alternative strategy is to target GC systems at closer distances. For example, NGC~253, a nearby spiral galaxy at a distance comparable to M81 ($\approx 3.5~\mathrm{Mpc}$), hosts approximately 100 GCs \citep{Cantiello2018}, making it a promising target for future searches.

\section*{Acknowledgements}
\rev{We would like to express our deepest appreciation to the anonymous reviewer for the comprehensive and thoughtful review of our manuscript.}
S.H., C.F., M.B., L.Z., E.C. and M.C. are supported by the Australian Research Council (ARC) Centre of Excellence (CoE) for Gravitational Wave Discovery (OzGrav) project numbers CE170100004 and CE230100016.
S.H. acknowledges the Astronomical Society of Australia (ASA) and the ANU Vice-Chancellor's HDR Travel Grants for travel support to the conference where this work was presented. 
M.C. acknowledges support of an Australian Research Council Discovery Early Career Research Award (project number DE220100819) funded by the Australian Government. T.H. acknowledges the support from the National Science and Technology Council (NSTC) of Taiwan through grants 113-2112-M-005-009-MY3, 113-2123-M-001-008, 111-2112-M-005-018-MY3, and the Ministry of Education of Taiwan through a grant 113RD109. T.G. acknowledges the support of the National Science and Technology Council of Taiwan through grants 113-2112-M-007-006, 113-2927-I-007-501, and 113-2123-M-001-008.
We thank Yunpeng Men, Vivek Venkatraman Krishnan and Ewan Barr for their technical support on \textsc{TransientX} and Joscha N. Jahns-Schindler for helpful discussions during this study.
FAST is a Chinese national mega-science facility, operated by the National Astronomical Observatories of Chinese Academy of Sciences (NAOC) \citep{2019SCPMA..6259502J, Jiang2020, 2020Innov...100053Q}.
This work used the OzSTAR national facility at Swinburne University of Technology. OzSTAR is funded by Swinburne University of Technology and the National Collaborative Research Infrastructure Strategy (NCRIS). 

%%%%%%%%%%%%%%%%%%%%%%%%%%%%%%%%%%%%%%%%%%%%%%%%%%
\section*{Data Availability}
The data underlying this article has been released from FAST Operation and Development Center and is publicly available under the project ID: PT2023\_0173.



\bibliographystyle{mnras}
\bibliography{GCFRB} % if your bibtex file is called example.bib

%\begin{thebibliography}{99}
%\end{thebibliography}

%%%%%%%%%%%%%%%%%%%%%%%%%%%%%%%%%%%%%%%%%%%%%%%%%%

%%%%%%%%%%%%%%%%% APPENDICES %%%%%%%%%%%%%%%%%%%%%
\appendix
\section{Dedispersion plan (ddplan)} \label{M49appendix}
To optimise sensitivity across the searched DM range, we generated a dedispersion plan using the \textsc{ddplan} utility from the PulsaR Exploration and Search TOolkit (\textsc{PRESTO}) software suite \citep{Ransom2011}. The plan was constructed for a DM range of 0--5000\,pc\,cm$^{-3}$ adopted in Section~\ref{sec:M49results}.
\begin{figure}
\centering
	\includegraphics[angle=270, width=250pt]{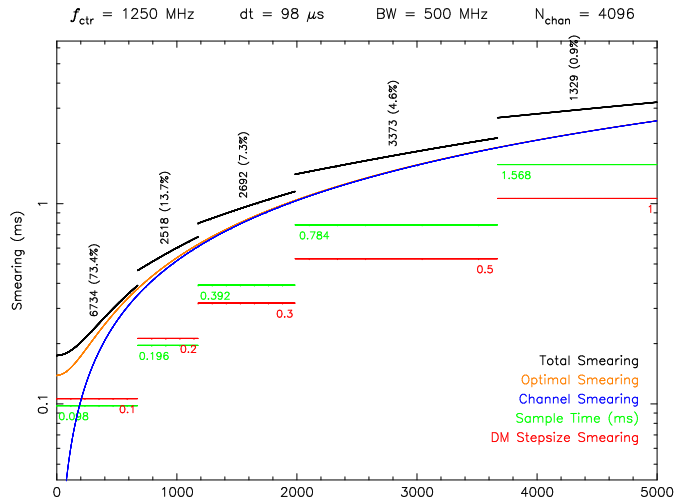}
    \caption{Smearing vs DM relation plot of a DDplan over the DM of 0--5000 pc cm$^{-3}$ produced by \textsc{PRESTO}. The numbers of DM trials are plotted as black broken lines and annotated above. The optimal smearing, channel smearing, sample time and DM stepsize smearing are plotted as orange, blue, green and red, respectively. The central frequency $f_{\text{ctr}}$, time resolution dt, bandwidth BW and number of frequency channels N$_{\text{chan}}$ are at the top.}
    \label{DDplan_dm5000_4k_98us.eps}
\end{figure}

\section{Examples of RFI events} \label{M49appendix2}

\begin{figure}
\centering
	\includegraphics[width=\linewidth]{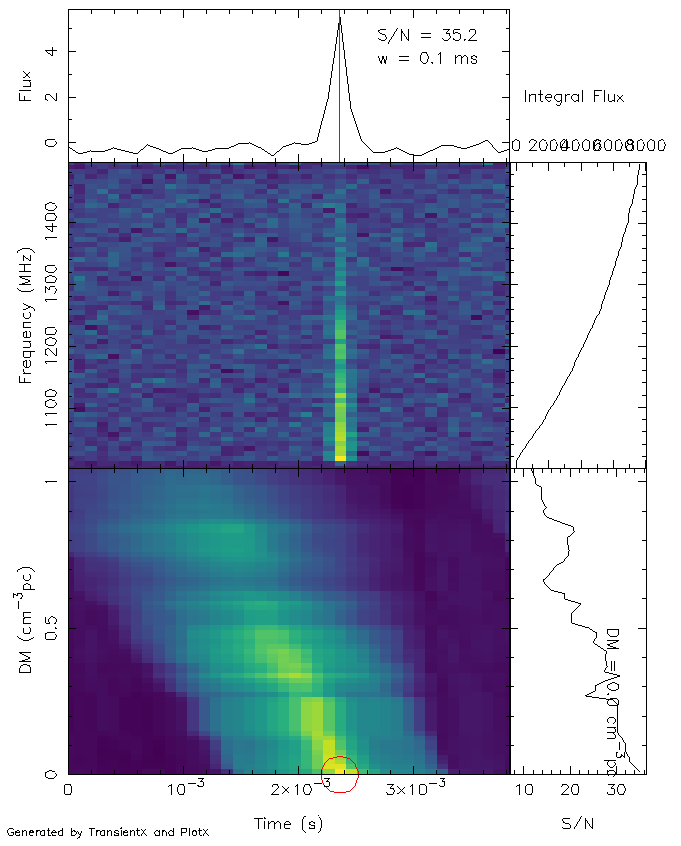}
    \caption{Representative examples of radio-frequency interference (RFI) events identified during the FAST single-pulse search. Each panel shows the frequency--time waterfall plot (top), the corresponding dedispersed pulse profile (middle), and the signal-to-noise ratio ($\mathrm{S/N}$) as a function of trial dispersion measure (bottom). The top panel shows a candidate with $\mathrm{DM}=0\,\mathrm{pc\,cm^{-3}}$. The bottom panel shows a long-duration RFI event.}
    \label{fig:RFI1}
\end{figure}

\begin{figure}
\centering
    \includegraphics[width=\linewidth]{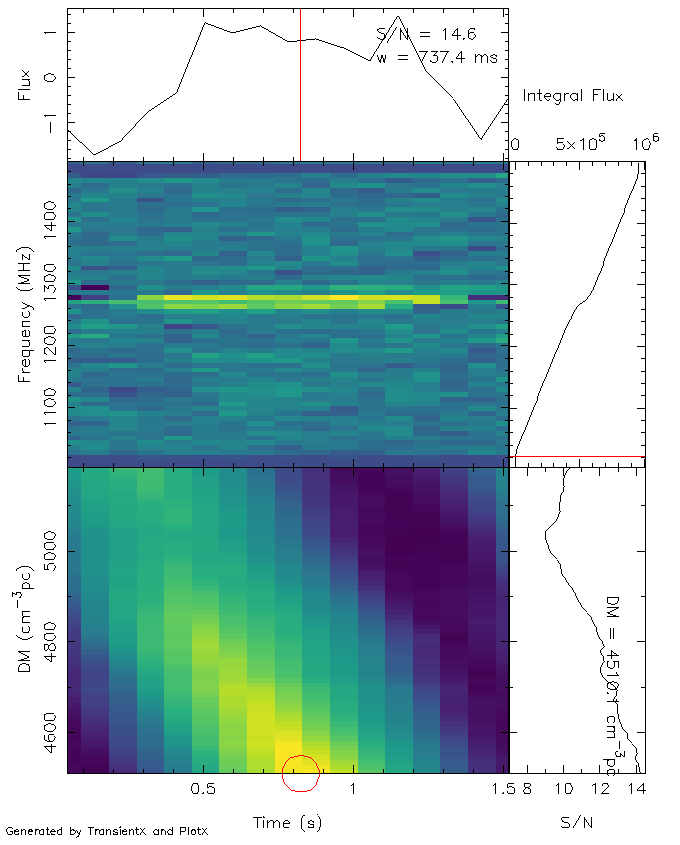}
    \includegraphics[width=\linewidth]{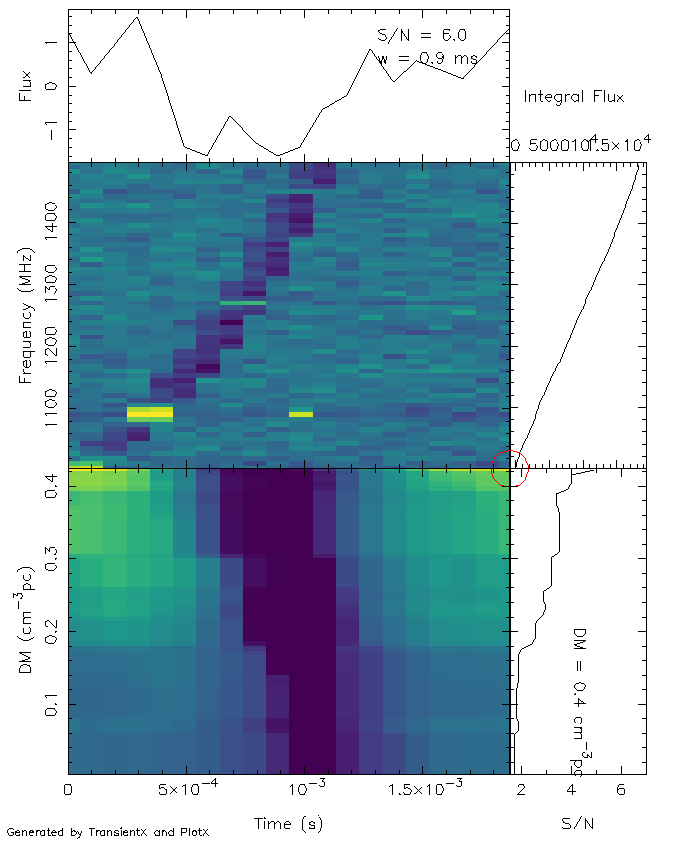}
    \caption{Representative examples of confirmed RFI events (continued), shown in the same format as Fig.~\ref{fig:RFI1}. The panel shows a short-duration RFI event.}
    \label{fig:RFI2}
\end{figure}

During the single-pulse search, a substantial fraction of candidates were attributed to radio-frequency interference (RFI). These events were identified based on their low DM values, narrow-band structure, non-dispersive frequency evolution, or atypical temporal morphologies. Representative examples are shown in Figures~\ref{fig:RFI1} and~\ref{fig:RFI2}. These examples illustrate the primary classes of RFI encountered in the dataset and demonstrate the criteria used for candidate rejection.

%%%%%%%%%%%%%%%%%%%%%%%%%%%%%%%%%%%%%%%%%%%%%%%%%%

% Don't change these lines
\bsp	% typesetting comment
\label{lastpage}
%\end{CJK*}
\end{document}